\documentclass[pre,twocolumn,aps,eqsecnum]{revtex4}
\usepackage{amssymb,amstext,amsmath,physics}
\usepackage{graphicx,diagbox}
\begin{document}

\title{Theory of transition from brittle to ductile fracture} 
\author{K.C. Le$^{1,2}$, H. Jeong$^3$, T.M. Tran$^4$} 
\affiliation{$^1$Division of Computational Mathematics and Engineering, Institute for Computational Science, Ton Duc Thang University, Ho Chi Minh City, Vietnam
\\
$^2$Faculty of Civil Engineering, Ton Duc Thang University, Ho Chi Minh City, Vietnam
\\
$^3$Lehrstuhl f\"ur Mechanik - Materialtheorie, Ruhr-Universit\"at Bochum, Bochum, Germany
\\
$^4$Department of Mechanical Engineering, Vietnamese German University, Binh Duong, Vietnam}

\date{\today}

\begin{abstract}
In this paper, two improvements to the theory of transition from brittle to ductile fracture developed by Langer are proposed. First, considering the drastic temperature rise near the crack tip, the temperature dependence of the shear modulus is included to better quantify the thermally sensitive dislocation entanglement. Second, the parameters of the improved theory are identified by the large scale least squares method. The comparison between the fracture toughness predicted by the theory and the values obtained in Gumbsch's experiments for tungsten at different temperatures shows good agreement.    
\end{abstract}

\maketitle

\section{Introduction}
\label{Intro} 

The transition from brittle to ductile fracture depends crucially on the rate of dislocation nucleation and multiplication under the elevated stress and temperature near the crack tip. Therefore, to develop a predictive theory for this transition, one must have a correct kinetics of dislocation depinning. Such a kinetics of dislocation depinning was first established in the framework of the thermodynamic dislocation theory (TDT), originally proposed by Langer, Bouchbinder and Lookmann \cite{LBL-10} and further developed in \cite{JSL-15,JSL-16,JSL-17,JSL-17a,Le17,Le18,Le20,LL20,LLT20,LDLG20}. The TDT  is based on two unconventional ideas. The first states that, under non-equilibrium conditions, the slow configurational degrees of freedom associated with the chaotic motion of dislocations are characterized by an effective disorder temperature that differs from the ordinary temperature caused by the much faster vibrations of atoms in the crystal lattice. Both temperatures are thermodynamically well-defined quantities whose equations of motion govern the irreversible behavior of these subsystems. The second basic idea is that dislocation entanglement is the overwhelming cause of resistance to deformation in crystals and that the dislocation depinning, determined by the double exponential formula, is strongly sensitive to small changes in stress and temperature. These two ideas have led to successful predictive theories of strain hardening \cite{LBL-10,JSL-15,LLT20,LDLG20}, steady-state stresses over wide ranges of temperatures and strain rates \cite{Le20,LL20},  thermal softening during plastic deformation \cite{Le17}, yielding transitions between elastic and plastic responses \cite{JSL-16,JSL-17a,LLT20}, and shear banding instabilities \cite{JSL-17,Le18}. 

To apply the TDT to the transition from brittle to ductile fracture, we need to analyze the local stress field near the crack tip. If the stress at the crack tip (called for short tip stress) increases fast enough, or if the plastic strain rate due to thermally activated dislocation depinning is sufficiently small, then this local tip stress caused by the prevailing elastic deformation quickly reaches a critical value at a relatively low stress intensity factor that triggers unstable crack growth or rapid crack propagation. This is a brittle fracture mode. On the other hand, when plastic deformation predominates, the crack tip is shielded by a plastic zone that prevents the tip stress from growing rapidly. The crack tip loses its shielding when the far-field stress exceeds a critical value that causes the unstably expanding plastic zone. At this point, the tip stress can quickly reach the breaking stress, resulting in crack growth, but at a much higher stress intensity factor and consequently a much higher fracture toughness. This is the ductile fracture mode. The equations governing the evolution of crack tip stress, crack tip curvature, dislocation density, and temperature as functions of stress intensity factor were recently proposed by Langer \cite{JSL-21}. With the ad hoc parameters chosen to be physically reasonable, he was able to simulate fracture toughness as a function of temperature and predict the transition from brittle to ductile fracture. Comparison with the fracture toughness as a function of temperature measured for tungsten in Gumbsch's experiments \cite{Gumbsch98,Gumbsch03} shows relatively good agreement.

The aim of this paper is to propose an improved theory of brittle-to-ductile transition (BDT) under the assumption of small strains. Similar to the theory developed by Langer in \cite{JSL-21}, we neglect the anisotropy effect and assume the elastic isotropy. However, compared to \cite{JSL-21}, there are two essential improvements. First, the temperature rises sharply near the crack tip and can reach several hundred or even thousand degrees near the ductile instability. The first experimental evidence of a temperature rise of about $130\,$K near the tip of a crack propagating at $10\,$m/s in steel was observed by Weichert and Sch\"onert \cite{WS78}. Another experimental measurement of the temperature rise of about $600\,$K in steel during adiabatic shear banding, at the end of which a crack forms, was made by Marchand and Duffy in \cite{MD88}. Light emission produced by rapid cracking in brittle materials has also been interpreted as thermal radiation at a temperature greater than $1000\,$K \cite{WS78}. Therefore, the dependence of the shear modulus and consequently of the Taylor stress on temperature must be taken into account to better quantify the thermally sensitive dislocation entanglement. Second, and most importantly, all material parameters otherwise chosen ad-hoc in \cite{JSL-21} must be replaced by those identified by the large scale least squares method \cite{LT17,Le17,LLT20,LDLG20}. The result is that the theoretical prediction based on this improved theory fits Gumbsch's experimental data much better than that of \cite{JSL-21}. 

The paper is structured as follows. Immediately after this short introduction, we start with the derivation of the equations of motion that will be used here. Similar to Langer's theory, these equations should be derived separately for two different cases: predeformed and non-predeformed crystals with an initial notch, which will be done in Sections~\ref{EOMP} and \ref{EOMN}, respectively. Our focus is on the physical meaning of the various parameters that occur in them. We discuss which of these parameters can be expected to be material-specific constants that are independent of temperature and loading rate, and thus are among the key parameters of the theory. In Section~\ref{NI}, we develop the large scale least squares method to determine the material parameters and compare the results of numerical simulations of toughness-temperature curves with the experimental data of Gumbsch {\it et al.} \cite{Gumbsch98}. We conclude in Section~\ref{CONCLUSIONS} with some remarks on the significance of these calculations.
 
\section{Equations of Motion: Predeformed Crystals}
\label{EOMP}
 
For simplicity, let us assume that a single crystal plate containing an initial notch deforms in plane strain mode I under a time-dependent remote tensile stress $\sigma_\infty(t)$. This idealized setting corresponds approximately to the three-point bending tests carried out by Gumbsch {\it et al.} \cite{Gumbsch98,Gumbsch03}. We first consider the case of a predeformed crystal, where a plastic zone already exists near the notch tip before the stress $\sigma_\infty(t)$ is applied. The mid-plane of the plate occupies an area in the $(x,y)$-plane, while the notch projection has the shape of an ellipse centered at the origin of the coordinate system, with a large semi-axis $W(1+m)$ in the $x$-direction and a much smaller semi-axis $W(1-m)$ in the $y$-direction ($m= 1-2\epsilon$, where $\epsilon$ is a small parameter). It is assumed that the elliptical shape does not change during the deformation of the notch into a crack. However, the curvature of the notch tip at $x=W(1+m)$, $k_\text{tip}$, may evolve during the loading process. Using the hypo-elasto-plasticity combined with TDT, Langer \cite{JSL-21} derived the following equation for the dimensionless curvature $\kappa =k_\text{tip} d_\text{tip}$, with $d_\text{tip}$ being the initial tip radius,
\begin{equation}
\label{kdot}
\frac{\dot{\kappa}}{\kappa }=\frac{(\bar{\nu}-1)^2}{3} \Bigl( \frac{\dot{s}_0}{2\mu} +D^{pl}_0\Bigr) +\Bigl( \frac{2\bar{\nu}-1}{\bar{\nu}^2}\Bigr) \Bigl( \frac{\dot{\sigma}_\infty }{2 \mu \epsilon}\Bigr).
\end{equation}
Here, the dot above a letter denotes the time derivative, $\bar{\nu}$ is function of the curvature and the yield stress characterizing the width of the plastic zone to be defined later, $s_0$ denotes the tip stress, $\mu$ is the shear modulus, and $D^{pl}_0$ corresponds to the plastic strain rate. The latter is given in the form \cite{LBL-10}
\begin{equation}
\label{q}
D^{pl}_0=\frac{b\sqrt{\rho}}{\tau_0} \exp \Bigl[ -\frac{T_P}{T} e^{-s_0/(\mu_T b \sqrt{\rho})} \Bigr] ,
\end{equation}
where $\tau_0$ is a microscopic timescale inversely proportional to the Debye frequency, $b$ is the Burgers vector, $s_0$ is the tip stress, $\rho$ is the dislocation density at the crack tip, $T_P$ is the pinning energy (in the temperature unit), while $\mu_T$ is a reduced shear modulus, so $\sigma_T=\mu_Tb\sqrt{\rho}$ gives the Taylor stress. The equation for the tip stress, derived by using a circular approximation as in \cite{JSL-20}, reads
\begin{equation}
\label{s0dot}
\frac{\dot{s}_0}{2\mu} =-D^{pl}_0+\Bigl( \frac{\dot{\sigma}_\infty }{2 \mu \epsilon}\Bigr) \frac{\Lambda(\bar{\nu})}{\bar{\nu}^3},
\end{equation}
where
\begin{displaymath}
\Lambda(\bar{\nu})=\frac{(\bar{\nu}-1)^2}{4\bar{\nu}\ln \bar{\nu}-(3+\bar{\nu})(\bar{\nu}-1)}.
\end{displaymath}
Note that function $\Lambda(\bar{\nu})$ has a singular point $\bar{\nu}_c\approx 5.116$, which plays a crucial role in predicting the instability of the plastic zone, as will be seen later on.

Langer assumed in \cite{JSL-21} that both $\mu$ and $\mu_T$ in Eqs.~\eqref{kdot}-\eqref{s0dot} are independent of temperature $T$. However, as mentioned in the introduction, the temperature changes drastically near the crack tip, especially during the unstable expansion of the plastic zone. Therefore, in order to better quantify the thermally sensitive dislocation entanglement, we take into account the dependence of $\mu$ on the temperature according to 
\begin{equation}
\label{muT}
\mu(T)=\mu_0 f(T), \quad f(T)=1-\frac{D/\mu_0}{e^{T_1/T}-1},
\end{equation}
where $\mu_0$, $D$, and $T_1$ are material constants (see Section IV for their values for tungsten). Note that Eq.~\eqref{muT} works well up to melting temperature \cite{VARSHNI-70}. Likewise, we assume that $\mu_T(T)=\mu_{T0}f(T)$. If we introduce the rescaled quantities
\begin{displaymath}
\bar{\rho}=a^2\rho, \quad \bar{s}_0=\frac{s_0}{\bar{\mu}_{T0}},\quad \bar{\mu}_{T}=\frac{b}{a}\mu_{T}=\bar{\mu}_{T0}f(T), 
\end{displaymath} 
where $a$ is a minimum spacing between dislocations, then the plastic strain rate can be rewritten as follows
\begin{equation}
\label{qb}
D^{pl}_0=\frac{q}{\tau_{pl}}, \quad q(\bar{s}_0,\bar{\rho},T)=\sqrt{\bar{\rho }} \exp \Bigl[ -\frac{T_P}{T} e^{-\bar{s}_0/(f(T)\sqrt{\bar{\rho}})} \Bigr],
\end{equation}
with $\tau_{pl}=a\tau_0/b$. 

To convert the equations \eqref{kdot} and \eqref{s0dot} into a form suitable for numerical simulation, we introduce the dimensionless stress intensity factor
\begin{equation}
\label{psi}
\psi=\frac{\sigma_\infty }{\bar{\mu}_{T}\epsilon \sqrt{\kappa}},
\end{equation}
and define
\begin{equation}
\label{psidot}
\dot{\psi}=\frac{\dot{\sigma}_\infty }{\bar{\mu}_{T}\epsilon \sqrt{\kappa}}\equiv \frac{1}{\tau_{ex}},\quad \xi\equiv \frac{\tau_{ex}}{\tau_{pl}},
\end{equation}
with $\xi$ being the inverse dimensionless loading rate. Note that the derivative of function $f(T)$ is negligibly small \cite{VARSHNI-70}, so that in the derivative of $\psi$ with respect to time $\bar{\mu}_T$ and $\kappa$ can be treated as constants. We assume that the loading rate represented by $\dot{\psi}$ is constant so that the time derivative can be replace by the derivative with respect to $\psi$ according to
\begin{displaymath}
\dv{t}=\tau_{ex}\dv{\psi}.
\end{displaymath} 
In this way the equations \eqref{kdot} and \eqref{s0dot} can be rewritten in the form
\begin{equation}
\label{kpsi}
\frac{1}{\kappa^{3/2}}\dv{\kappa}{\psi}=c_0\Bigl[ \frac{(\bar{\nu}-1)^2}{3}\frac{\Lambda(\bar{\nu})}{\bar{\nu}^3} + \frac{2\bar{\nu}-1}{\bar{\nu}^2} \Bigr] ,
\end{equation}
and
\begin{equation}
\label{spsi}
\dv{\bar{s}_0}{\psi}=-\frac{\xi}{c_0}q(\bar{s}_0,\bar{\rho},T)+\sqrt{\kappa}\frac{\Lambda(\bar{\nu})}{\bar{\nu}^3},
\end{equation}
where $c_0=\bar{\mu}_{T}/2\mu$. To define function $\bar{\nu}$ in these equations we must find the yield stress. It is obvious that the plastic yielding occurs when the elastic strain rate $s_0/2\mu$ becomes negligibly small. By inverting \eqref{qb} where the total strain rate stands on the left-hand side and using similar deliberations as in \cite{JSL-21}, we get the dimensionless yield stress in the form
\begin{equation}
\label{syb}
\bar{s}_y=f(T)\sqrt{\bar{\rho}}\Bigl\{ \ln \Bigl( \frac{T_P}{T}\Bigr) -\ln \Bigl[ \ln \Bigl( \frac{10\xi }{c_0 \ln(T_P/T)}\Bigr) \Bigr] \Bigr\} .
\end{equation}
Then $\bar{\nu}$, as function of $\bar{s}_y$, should be defined as follows
\begin{equation}
\label{nub}
\bar{\nu}(y)=\begin{cases}
   y^{1/3}   & \text{if $y>1$}, \\
   1   & \text{otherwise},
\end{cases}
\end{equation}
with $y=\psi \sqrt{\kappa}/\bar{s}_y$.

To close the system of equations we need to write down the equations for the dislocation density $\bar{\rho}$ and the equation for the kinetic-vibrational temperature $T$. Both equations can be proposed in accordance with the principles of irreversible thermodynamics. The equation for $\bar{\rho}$ remains unchanged compared to that in \cite{JSL-21}
\begin{equation}
\label{rhopsi}
\dv{\bar{\rho}}{\psi}=A \xi q(\bar{s}_0,\bar{\rho},T) \bar{s}_0(\psi) \Bigl( 1-\frac{\bar{\rho}}{\bar{\rho}_\infty }\Bigr) .
\end{equation}
It describes the evolution of dislocations in the simplified situation of a predeformed crystal when the configurational entropy is already close to the highest possible value. The interpretation of \eqref{rhopsi} is simple: The rate at which dislocations are formed is proportional to the plastic power and the detailed-balance factor $1-\bar{\rho}/\bar{\rho}_\infty $, which accounts for dislocation annihilation by requiring $\bar{\rho}$ to approach its saturated value $\bar{\rho}_\infty $. We assume that $A$ is independent of temperature and the inverse loading rate $\xi$. The remaining equation for $T$ reads
\begin{equation}
\label{tpsi}
\dv{T}{\psi}=C(T)\xi q(\bar{s}_0,\bar{\rho},T) \bar{s}_0(\psi).
\end{equation}
The term on the right-hand side corresponds to the plastic power that dissipates into heat, with $C(T)=C_0\exp (T_A/T)$ the Taylor-Quinney factor. The exponential dependence of $C$ on $T_A/T$ is indirectly justified by comparison with the experimental stress-strain curves in the problems of thermal softening \cite{Le17} and adiabatic shear banding \cite{Le18}. Certainly, the heat exchange between the crack tip and the surrounding material could also be considered. Since this leads to the models with heat conduction which are non-local, we will neglect this heat exchange for simplicity.

\section{Equations of Motion: Non-predeformed Crystals}
\label{EOMN}  

For non-predeformed crystals with an initial notch, the situation changes drastically. In this case, the redundant dislocations of opposite sign are initially in the pinned state and are therefore inactive in an early stage of deformation. When the load is applied and the tip stress is not yet high enough to overcome the pinning energy barrier but sufficiently high to emit dislocations at the notch tip, non-redundant dislocations of the same sign (geometrically necessary dislocations) form at the notch tip and then quickly move away under the drag force from there, changing the curvature of the notch \cite{RT-74,Rice-92}. This type of dislocations is called by Langer ``dragged'' dislocations, their density is denoted by $\rho_D$. Since this type of dislocations is recognized by us as non-redundant dislocations, we change the notation for their density to $\rho_G$. The depinning of redundant dislocations, called by Langer ``entangled'' dislocations, occurs at a later stage of deformation, when the tip stress is high enough to overcome the pinning energy barrier, leading to separation of dislocations of opposite sign. Since this is the thermally assisted collective dislocation generation \cite{KT-73,KPV-94}, their density, which we denote by the same letter $\rho$ as in the previous case, becomes dominant rapidly thereafter. With this analysis in mind, we now turn to the equations of motion at the notch tip, whose derivation is essentially the same as Langer's, although there are some differences in detail. Since the emission of non-redundant dislocations increases the tip curvature as mentioned above (see \cite{RT-74,Rice-92}), Eq.~\ref{kpsi} should change to
\begin{equation}
\label{kpsin}
\frac{1}{\kappa^{3/2}}\dv{\kappa}{\psi}=c_0\Bigl[ \frac{(\bar{\nu}-1)^2}{3}\frac{\Lambda(\bar{\nu})}{\bar{\nu}^3} + \frac{2\bar{\nu}-1}{\bar{\nu}^2} \Bigr] +\frac{\bar{\rho}_G\psi}{\xi \eta(T)}.
\end{equation}
The last term on the right-hand side of this equation is the contribution of the non-redundant dislocations to the change of the tip curvature, with $\bar{\rho}_G$ being their dimensionless density in unit $a^{-2}$ and $\eta (T)=\eta_0 e^{-T_D/T}$ a thermal activation factor (Gumbsch's scaling law \cite{Gumbsch03}). Likewise, the presence of non-redundant dislocations also causes the change of tip stress, so the equation for $\bar{s}_0$ becomes
\begin{equation}
\label{spsin}
\dv{\bar{s}_0}{\psi}=-\frac{\xi}{c_0}q(\bar{s}_0,\bar{\rho},T)+\sqrt{\kappa}\frac{\Lambda(\bar{\nu})}{\bar{\nu}^3}+\frac{\bar{\rho}_G\psi^2 \kappa}{\xi \eta(T)}.
\end{equation}
The last term in Eq.~\eqref{spsin} is the contribution of the non-redundant dislocation to the change in tip stress. The presence of the $\xi$-factor in the denominators of the last terms in Eqs.~\eqref{kpsin} and \eqref{spsin} means that the increase in tip curvature and tip stress is proportional to the loading rate. This can be explained by the fact that the dislocation emission from the tip occurs in times much shorter than the characteristic time for curvature and stress growth. Therefore, the generalized Orowan formula, which contains the number of dislocations in a square whose side length is proportional to the loading rate, leads to the $\xi$-factor in the denominators of these terms \cite{JSL-21}. Eqs.~\eqref{kpsin} and \eqref{spsin} are identical in form to Eqs.~(5.9) and (5.10) in \cite{JSL-21}, except that functions $q(\bar{s}_0,\bar{\rho},T)$ and $\bar{s}_y$ are now given by \eqref{qb} and \eqref{syb}, respectively.

Since we have now two families of dislocations, their densities must obey two separate evolution equations. The equation for $\bar{\rho}$ remains unchanged except that the pre-factor $A \xi $ is no longer a linear function of $\xi$ as in the previous case. Assuming that this pre-factor is a function of $\xi$, we write the equation for $\bar{\rho}$ in a slightly different form as \cite{JSL-21}
\begin{equation}
\label{rhopsin}
\dv{\bar{\rho}}{\psi}=A(\xi) q(\bar{s}_0,\bar{\rho},T) \bar{s}_0(\psi) \Bigl( 1-\frac{\bar{\rho}}{\bar{\rho}_\infty }\Bigr) .
\end{equation}
Our proposition for $A(\xi)$ also differs significantly from that of \cite{JSL-21}: We now assume that $A(\xi)=A_1 \exp(A_2 \xi)$. Thus, as $\xi$ goes to zero (very fast loading) function $A(\xi)$ approaches the constant value $A_1$. Regarding $\rho_G$, the density of non-redundant dislocations: This quantity must be expressed by the curl of the nonuniform plastic distortion satisfying the equilibrium of micro-forces acting on them, which is actually a partial differential equation \cite{LT-16,L-18}. We simplify the situation by posing the following evolution equation for the local tip density $\bar{\rho}_G$
\begin{equation}
\label{rhogpsin}
\dv{\bar{\rho}_G}{\psi}=A_n \frac{\psi^2 \bar{\rho}_G\kappa (\psi)}{\xi \eta(T)} \Bigl( 1-\frac{\bar{\rho}_G}{\bar{\rho}_c }\Bigr) .
\end{equation}
The detailed-balance factor on the right-hand side of this equation contains a value $\bar{\rho}_c =t_n \bar{\rho}_\infty $ that is much smaller than $\bar{\rho}_\infty $. This means that most of the non-redundant dislocations move away from the crack tip, so the local density of these dislocations accounts for only a small fraction of the total density of dislocations near the crack tip in the later stage of deformation. Finally, the equation for $T$ remains the same as \eqref{tpsi} except that the Taylor-Quinney factor $C(T)$ must be replaced by $C_n(T)=C_{0n}\exp (T_{An}/T)$. The replacement of the function $C(T)$ by $C_n(T)$ can be explained by the fact that the dissipation into heat caused by dislocations in this case is a much longer process starting with a negligibly small dislocation density. 

\section{Data analysis}
\label{NI}

\begin{figure}[h]
\centering \includegraphics[width=.49\textwidth]{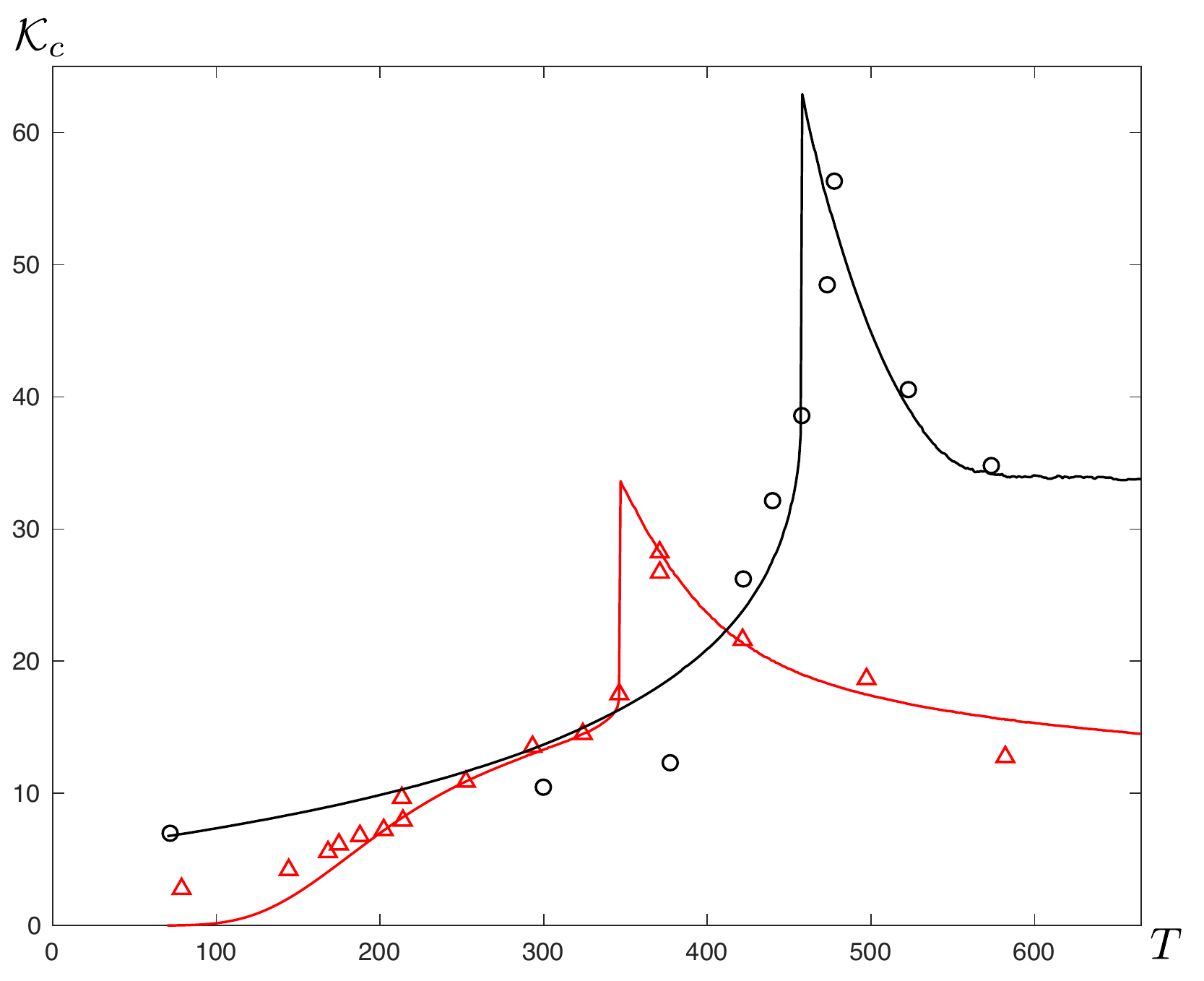} \caption{(Color online) Fracture toughness $\mathcal{K}_c$ (in unit $\text{MPa}\, \text{m}^{1/2}$) versus temperature (in unit K): (i) predeformed tungsten at loading rate $0.1\, \text{MPa}\, \text{m}^{1/2}\, \text{s}^{-1}$ (black curve, circles), (ii) non-predeformed tungsten at loading rates  $0.1\, \text{MPa}\, \text{m}^{1/2}\, \text{s}^{-1}$ (red curve, triangles).  The experimental points are taken from Gumbsch {\it et al.} \cite{Gumbsch98}} \label{fig:1}
\end{figure}

The experimental results of Gumbsch {\it et al.} \cite{Gumbsch98} alongside with our theoretical predictions based on the equations of motion \eqref{kpsi}, \eqref{spsi}, \eqref{rhopsi}, and \eqref{tpsi} in predeformed case, and \eqref{kpsin}, \eqref{spsin}, \eqref{rhopsin}, \eqref{rhogpsin}, and \eqref{tpsi} in non-predeformed case, are shown in Fig.~\ref{fig:1}.

In order to obtain the fracture toughness from the theory, the above systems of ODEs complemented by the initial conditions must first be solved. In the predeformed case the initial conditions are assumed in the form
\begin{equation}
\label{ini}
\kappa(0)=1,\, \bar{s}_0(0)=0,\, \bar{\rho}(0)=r_i \bar{\rho}_\infty,\, T(0)=T_0.
\end{equation}
In the non-predeformed case we pose
\begin{equation}
\label{inin}
\kappa(0)=1,\, \bar{s}_0(0)=0,\, \bar{\rho}(0)=\bar{\rho}_{i} ,\, \bar{\rho}_G(0)=\bar{\rho}_{Gi},\, T(0)=T_0.
\end{equation}
Having found the solution of the ODEs, we can determine the critical stress intensity factor $\psi_c$ at which the crack starts to grow as follows: In the predeformed case, it is the smallest root of the equation $\bar{s}_0(\psi_c)=\bar{s}_c$, while in the non-predeformed case it is the smallest root of the equation $\bar{s}_0(\psi_c)=\bar{s}_{cn}$, where $\bar{s}_{cn}$ is somewhat larger than $\bar{s}_{c}$. The explanation is rather simple: the predeformed crystal has already been subjected to strain hardening, which has created new defects or weakened existing ones, so $\bar{s}_c$ must be smaller than $\bar{s}_{cn}$. Finally, the fracture toughness is evaluated as $\mathcal{K}_c=\alpha \psi_c$. Note that the proportionality factor $\alpha$ depends only on instrumentation and not on sample preparation.

Thus, in order to simulate the theoretical toughness as shown in Fig.~\ref{fig:1}, we need numerical values for four system-specific parameters common in both cases: the pinning energy $T_P$ (in temperature unit), the moduli ratio $c_0$, the dimensionless saturated dislocation density $\bar{\rho}_\infty$, and the toughness factor $\alpha$. In predeformed case we need additionally four parameters: $A$, $C_0$, $T_A$, $\bar{s}_{c}$, while in the non-predeformed case there are nine additional parameters: $\eta_0$, $T_D$, $A_1$, $A_2$, $C_{0n}$, $T_{An}$, $A_n$, $\bar{s}_{cn}$, and $t_n$. Besides, to specify the initial conditions, we need $r_i$ in the predeformed case and $\bar{\rho}_i$ and $\bar{\rho}_{Gi}$ in the non-predeformed case. For the temperature-dependent shear modulus $\mu(T)$  of tungsten given by \eqref{muT} we take $\mu_0= 159.5\,$GPa, $D= 33.69$, $T_1=1217\,$K \cite{{DLRM-98}}. 

All the above parameters are present in Langer's theory \cite{JSL-21}. He has chosen most of them ad hoc, based on some physical deliberations. Some parameters like $A_1$, $A_2$, $\bar{s}_{cn}$ or $C_{0n}$ even vary from case to case. In this paper, we abandon this way of choosing parameters and try to identify them by the large scale least squares method. Let the unknown parameters be the components of a vector denoted by $\mathbf{P}$, which belongs to the multidimensional space of parameters. This vector is obtained from $N_e$ experimentally measured toughness-temperature curves (of $N_e$ different cases and loading rates) as follows. Assuming $\mathbf{P}$ is known, we integrate the system \eqref{kpsi}, \eqref{spsi}, \eqref{rhopsi} and \eqref{tpsi} or \eqref{kpsin}, \eqref{spsin}, \eqref{rhopsin}, \eqref{rhogpsin}, and \eqref{tpsi} depending on the cases and loading rates to find $N_e$ functions $\mathcal{K}_{ci}(T,\mathbf{P})$, $i=1,\ldots ,N_e$. Then we form the sum of squares
\begin{displaymath}
h(\mathbf{P})=\sum_{i=1}^{N_e}\sum_{j=1}^{N_i} (\mathcal{K}_{ci}(T_{ij},\mathbf{P})-\mathcal{K}^*_{ij})^2,
\end{displaymath}
where $(T_{ij}, \mathcal{K}^*_{ij})$, $i=1,\ldots ,N_e$, $j=1,\ldots N_i$ correspond to the temperatures and fracture toughnesses measured in the experiment at $N_e$ different cases and loading rates (the index $j$ goes from 1 to $N_i$, where $N_i$ is the number of selected points on the curve $i$). We then find the optimal $\mathbf{P}$ by minimizing the function $h(\mathbf{P})$ in the space of parameters subject to physically meaningful constraints. Langer's physical deliberations are still very valuable and can be used to obtain initial guesses and reasonable upper and lower bounds for the parameters. Since the above ODE-systems are stiff, we use the Matlab solver ode15s for their numerical integration. Regarding the minimization of the function $h(\mathbf{P})$: As we are aiming at the global minimum (least squares), the best numerical package for this is the Matlab GlobalSearch minimization. 

\begin{figure}[h]
\centering \includegraphics[width=.49\textwidth]{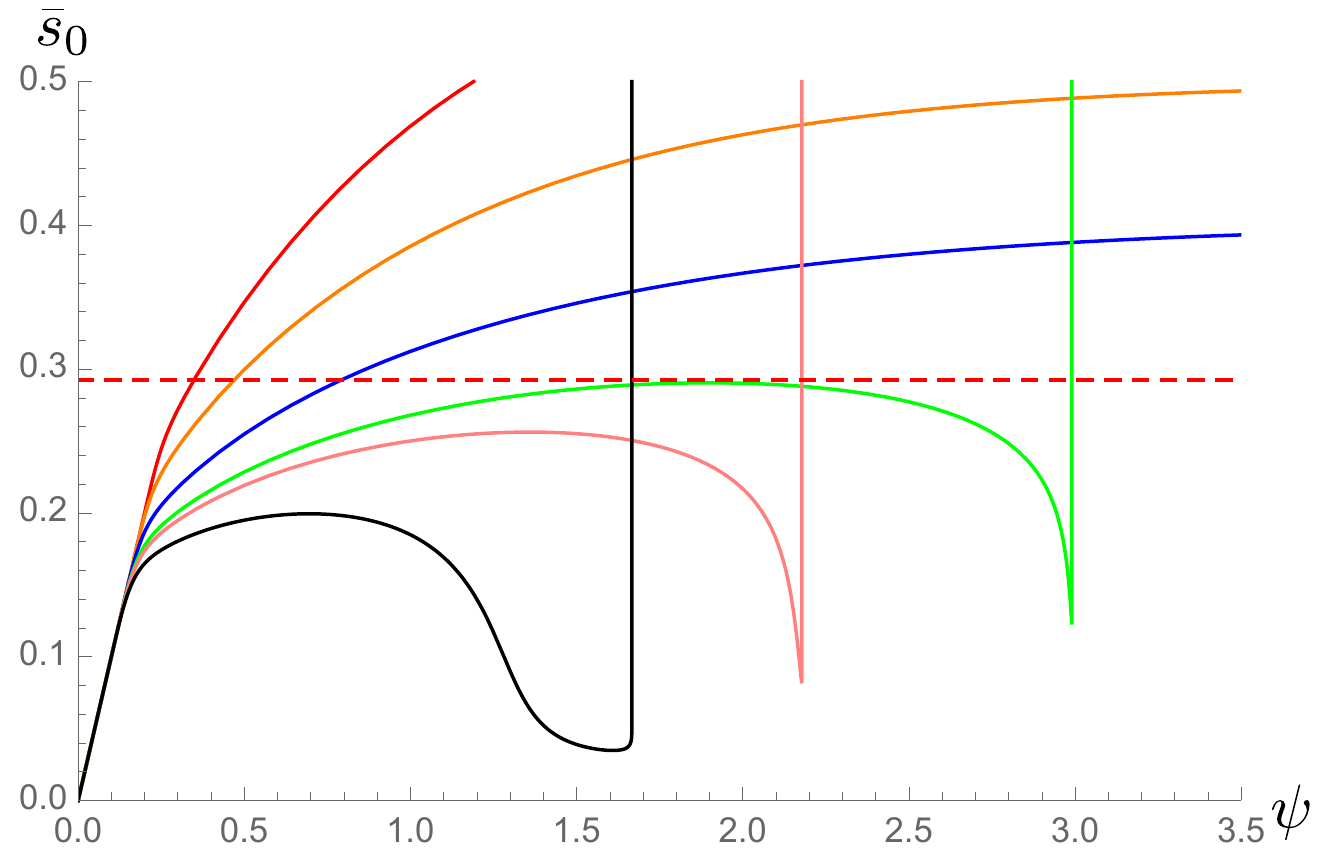} \caption{(Color online) Tip stresses $\bar{s}_0(\psi)$ representing the solution of the ODEs \eqref{kpsi}-\eqref{tpsi} for temperatures $T_0=100\,$K, $200\,$K, $350\,$K, $460\,$K, $500\,$K, $600\,$K, from top to bottom, plus a dashed line at the breaking stress $\bar{s}_c$ for predeformed tungsten at the loading rate $\xi=1$.}
\label{fig:2}
\end{figure}

One of the numerical challenges is the evaluation of fracture toughness in the context of finding the smallest root of the equation $\bar{s}_0(\psi_c)=\bar{s}_c$. As shown in Fig.~\ref{fig:2}, the behavior of the tip stress $\bar{s}_0(\psi)$ representing the solution of the ODEs \eqref{kpsi}-\eqref{tpsi} for predeformed crystal (at the loading rate $\xi=1$) changes drastically when $T$ exceeds a critical value $T_c\approx 460\,$K. At low temperatures ($T<T_c$), the plastic shielding is small and the tip stress is a monotonically increasing function of $\psi$ and shows a nearly elastic behavior, so it is straightforward to find the smallest root of the equation $\bar{s}_0(\psi_c)=\bar{s}_c$. For $T=T_c$ the tip stress, which is the green curve, touches the horizontal line $\bar{s}_0=\bar{s}_c$, while for $T>T_c$ the plastic shielding becomes so strong that the tip stress does not reach the horizontal line before the boundary layer experiences its thermal instability, leading to the singular behavior of $\bar{s}_0(\psi)$ indicated by a (nearly) vertical line (the pink and black curves). Thus, the root is very close to the singular point when function $\Lambda(\bar{\nu})$ goes to infinity, i.e. $\bar{\nu}_c\approx 5.116$. Despite some minor differences in detail, the behavior of $\bar{s}_0(\psi)$ representing the solution of the ODEs \eqref{kpsin}-\eqref{rhogpsin} and \eqref{tpsi} for non-predeformed crystals at the loading rate $\xi=1$ is similar, as shown in Fig.~3. In this case, the temperature of BDT is $T_c=360\,$K. Since the numerical integration stops when the $\psi$-step is smaller that some fixed value, it is not always possible to get $s_0(\psi)$ near the singular point reaching the horizontal line $s_0=s_c$ and to find the  intersection point. In order to do so we choose the  alternative event that $\bar{\nu}$ is larger than $5.115$. If either of the conditions $\bar{s}_0>s_c$ and $\bar{\nu}>5.115$ is satisfied, then the root finding is stopped and the last $\psi $ is output as $\psi_c$. 

	\begin{table}[h]
	\begin{center}
		\begin{tabular}{ | l | l | l | l | l |}
			\hline
			\diagbox[width=10em]{Theory}{Parameter} & $T_P\,$(K) & $c_0$  & $\bar{\rho}_\infty$ & $\alpha $  \\ \hline
			Our    & 36214 & $1.026\, 10^{-2}$  & $1.79\, 10^{-2}$   & $21.1$  \\ \hline
			Langer    & 36000  & 0.01  & $1.83\, 10^{-2}$  & 20  \\ 
 \hline
		\end{tabular}
		\caption{\label{tab:TDTParameters1} Common material parameters for tungsten.}
	\end{center}
	\end{table}
	
	\begin{table}[h]
	\begin{center}
		\begin{tabular}{ | l | l | l | l | l |}
			\hline
			 & $A$ & $C_0$  & $T_A\,$ (K) & $\bar{s}_c$  \\ \hline
			Our    & 11.58 & $5.6\, 10^7$ & $3283$   & $0.2934$  \\ \hline
			Langer    & 10 & $7\, 10^7$ & $3500$  & $0.31$ \\ 
 \hline
		\end{tabular}
		\caption{\label{tab:TDTParameters2} Parameters for predeformed tungsten.}
	\end{center}
	\end{table}
	
	\begin{table}[h]
	\begin{center}
		\begin{tabular}{|clc|clc|clc|clc|cl}
			\hline
			 $\eta_0$ & $T_D\,$(K)  & $A_1$ & $A_2$ & $C_{0n}$ & $T_{An}$ & $A_n$ & $\bar{s}_{cn}$ & $t_n$  \\ \hline
			0.876 & $1728$ & $7.993$ & $4.318$  & $7.7\, 10^7$ & $2912$ & $12.27$ & $0.473$ & $9.07\, 10^{-2}$ \\ \hline
			1.2 & $2200$ & -  & - & - & $3500$ & 10 & - & 0.1  \\ 
 \hline
		\end{tabular}
		\caption{\label{tab:TDTParameters2} Parameters for non-predeformed tungsten.}
	\end{center}
	\end{table}

With the developed large scale least squares method we could identify the optimal  parameters used in simulating the toughness-temperature curves shown in Fig.~\ref{fig:1}. The common parameters are listed in Table~I, those for predeformed tungsten in Table~II, and those for non-predeformed tungsten in Table~III. For comparison, we also show the values used by Langer in \cite{JSL-21} in the next rows of these tables, except for the parameters, which vary from case to case. Note that the value of $T_P$ is close to that found in \cite{LDLG20}. The value of $T_D$ is somewhat smaller than Gumbsch's scaling temperature $2200\,$K. This suggests that the change in curvature and growth of tip stress due to nucleation and emission of non-redundant dislocations are less sensitive to thermal fluctuations. Finally, the parameters related to the initial dislocation densities are: $r_i=0.145$, $\bar{\rho}_i=9.67\, 10^{-8}$, $\bar{\rho}_{Gi}=6.89\, 10^{-5}$.

\begin{figure}[h]
\centering \includegraphics[width=.49\textwidth]{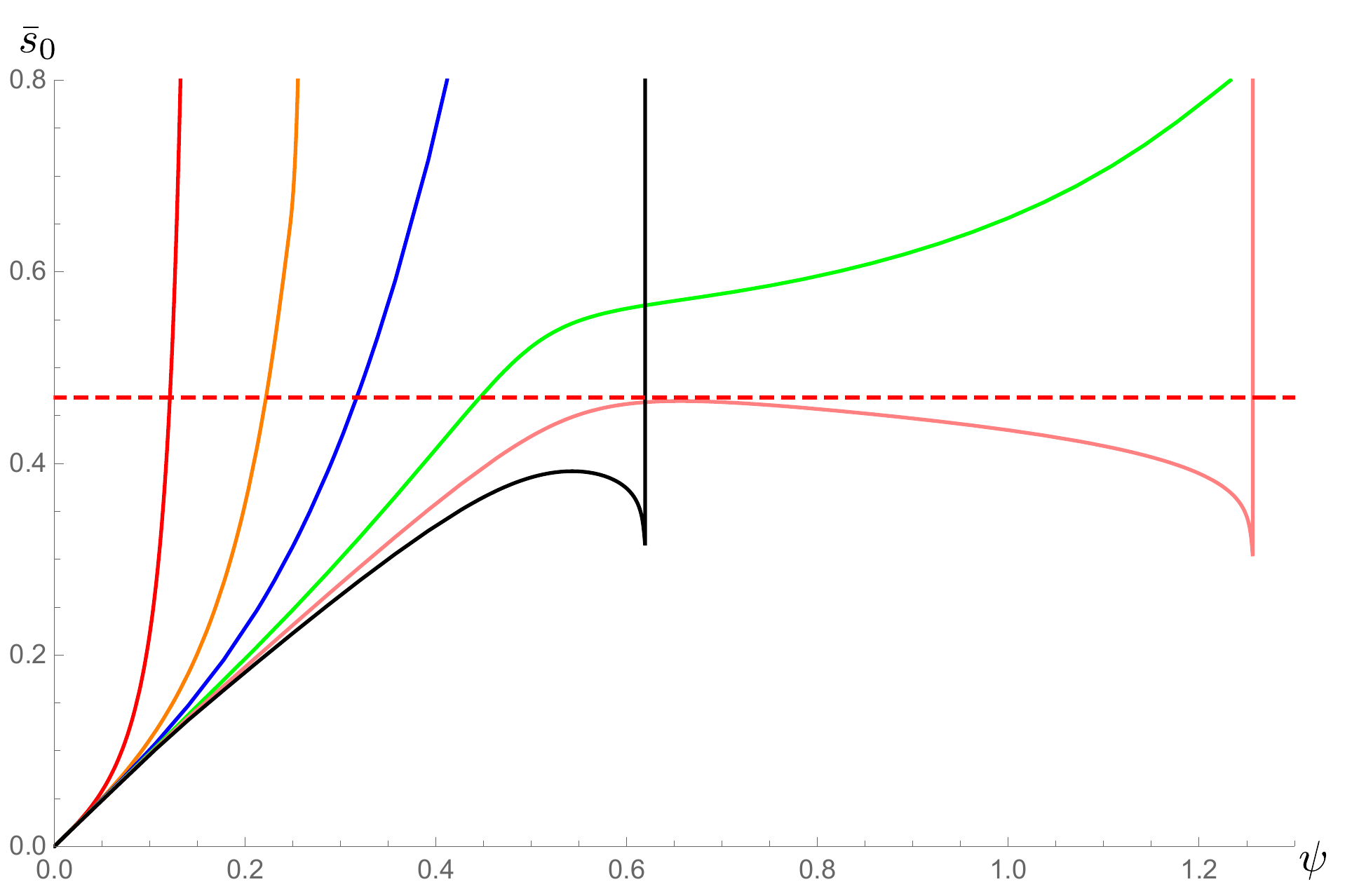} \caption{(Color online) Tip stresses $\bar{s}_0(\psi)$ representing the solution of the ODEs \eqref{kpsin}-\eqref{rhogpsin} and \eqref{tpsi} for temperatures $T_0=150\,$K, $175\,$K, $200\,$K, $250\,$K, $360\,$K, $500\,$K, from top to bottom, plus a dashed line at the breaking stress $\bar{s}_c$ for non-predeformed tungsten at the loading rate $\xi=1$.}
\label{fig:3}
\end{figure}

The toughness-temperature curves calculated according to the proposed theory with the above parameters (see Fig.~1) show a remarkable behavior: At low temperatures, the toughness is rather low and increases with increasing temperature. At temperature $T_c\approx 460\,$K (for predeformed tungsten) (and $T_c\approx 360\,$K for non-predeformed tungsten) the toughness experiences a sharp jump corresponding to the vertical line. When the temperature exceeds $T_c$, the toughness decreases with increasing temperature, but still remains higher than before BDT. Therefore, $T_c$ is considered to be the temperature of BDT. Note the good agreement between the toughness-temperature curves predicted by theory and the experimental points measured in \cite{Gumbsch98}. A noticeable deviation is observed only at two temperatures $T=300\,$K and $T= 377\,$K for predeformed tungsten and at lowest and highest temperatures for non-predeformed tungsten. Considering various  uncertainties in sample preparation and toughness measurement methods, the agreement seems satisfactory.

\begin{figure}[h]
\centering \includegraphics[width=.49\textwidth]{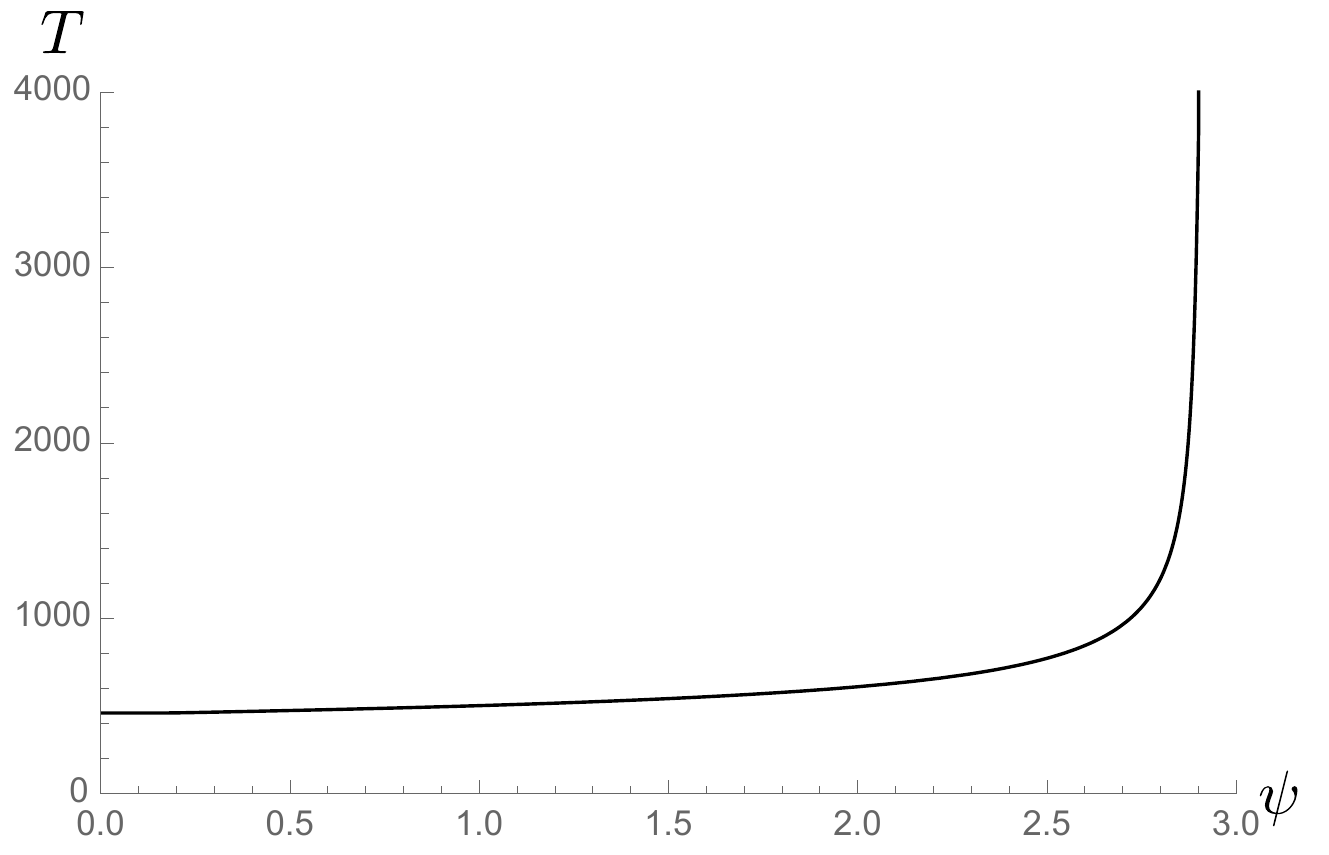} \caption{Tip temperature $T(\psi)$ for non-predeformed tungsten at the loading rate $\xi=1$ and at the remote temperature $460\,$K.}
\label{fig:4}
\end{figure}

Fig.~4 shows the change of the tip temperature for predeformed tungsten at loading rate $\xi=1$ and at remote temperature $460\,$K during the deformation process. We see that the temperature changes only slightly in the early stage of deformation. However, during the unstable expansion of the plastic zone, the temperature changes drastically. This is caused by the sharp increase in tip stress, which leads to a large plastic power, a main part of which is dissipated into heat (compare with the somewhat similar material instability during adiabatic shear banding, which leads to an increase in plastic strain rate and plastic power \cite{Le18}). This the reason why taking into account the temperature dependence of the shear modulus could improve the quantitative agreement with the experiment at the later stage of deformation.

\section{Concluding Remarks}
\label{CONCLUSIONS}

Based on the above analysis, we can conclude that taking into account the temperature dependence of the shear modulus as well as the large-scale least squares method for identifying the material parameters can significantly improve the agreement between theory and experiment. However, the theory could still be improved in many ways. For example, we neglect completely the change in configurational temperature in the equations of motion. This is perhaps acceptable for the case of predeformed crystals, but not fully justified for the case of non-predeformed crystals. In addition, the simplified theory does not account for the spatial distribution of stress and plastic distortion, or for heat transfer by conduction. Another nonlinear effect that could play an important role is the decrease of the tangential stiffness matrix, which could accelerate the material instability and lead to the formation of an adiabatic shear band in front of the crack tip. The nonlinear and nonlocal model, which analyzes the stress field and inhomogeneous plastic distortion in the plastic zone ahead of the crack tip, as well as heat conduction, will be discussed elsewhere in the future.

\begin{acknowledgments}

K.C. Le is grateful to J.S. Langer for helpful discussions. We would also like to thank the reviewers for various constructive suggestions. 

\end{acknowledgments}

\end{document}